\documentclass[journal]{IEEEtran}

\usepackage{amsmath,amssymb,amsfonts}
\usepackage{algorithmic}
\usepackage{graphicx}
\usepackage{textcomp}
\def\BibTeX{{\rm B\kern-.05em{\sc i\kern-.025em b}\kern-.08em
    T\kern-.1667em\lower.7ex\hbox{E}\kern-.125emX}}
\DeclareGraphicsExtensions{.pdf,.jpeg,.png}
\graphicspath{{figures/}}

\usepackage{xcolor}
\usepackage{color}
\usepackage[nolist]{acronym}
\usepackage[normalem]{ulem}
\usepackage{animate}

\usepackage{cite}
\usepackage{tikz}
\usepackage{pgfplots}

\newcommand\figwidth{8.9} % in cm
\newacro{MoM}[MoM]{method of moments}
\newacro{MOO}[MOO]{multiobjective optimization}
\newacro{CM}[CM]{characteristic mode}
\newacro{PEC}[PEC]{perfect electric conductor}
\newacro{PMC}[PMC]{perfect magnetic conductor}
\newacro{EP}[EP]{eigenvalue problem}
\newacro{GEP}[GEP]{generalized eigenvalue problem}
\newacro{EFIE}[EFIE]{electric field integral equation}
\newacro{SVD}[SVD]{singular value decomposition}
\newacro{RWG}[RWG]{Rao-Wilton-Glisson}
\newacro{EM}[EM]{electromagnetic}
\newacro{dof}[d-o-f]{\mbox{degrees-of-freedom}}
\newacro{TEM}[TEM]{transverse electromagnetic}
\newacro{TRL}[TRL]{transmission line}

\begin{document}

\title{Wave-Guiding Part of a Capacitor Paradox}
\author{Petr Ourednik, Lukas~Jelinek
%\thanks{Manuscript received November XX , 2017; revised November XX, 2017.}
%\thanks{This work was supported by the Department of electromagnetic field at CTU FEE Prague}
\thanks{P.~Ourednik and L.~Jelinek are with the Department of Electromagnetic Field, Faculty of Electrical Engineering, Czech Technical University in Prague, Technicka~2, 16627, Prague, Czech Republic
	(e-mail: \mbox{ouredpet@fel.cvut.cz}, \mbox{lukas.jelinek@fel.cvut.cz}).}
}

%%%%%%%%%%%%%%%%%%%%%%

%\markboth{Journal of \LaTeX\ Class Files,~Vol.~XX, No.~XX, November~2017}
%{Jelinek \MakeLowercase{\textit{et al.}}: Capacitor Paradox}
\maketitle

%\begin{abstract}
%XXXXX
%\end{abstract}

%%%%%%%%%%%%%%%%%%%%%%%%%%%%%%%%%%%%%%%%%%%%%%%%%%%%%%%%%%%%%%%%%%%%%%%%%%%%%%%%%%%%%%%%%%%%%%%%%%%%%%%%%%%%
\section{Introduction}
\label{Intro}

A capacitor paradox is an electromagnetic problem designed to show the limits of circuit theory which considers the connection of two ideal capacitors, one charged and the second discharged. A blind solution to this problem leads to the non-conservation of energy and, thus, to a paradox. A detailed and accurate explanation of the treatment of this problem can be found in~\cite{McDonald_capacitor_paradox}, including an extensive citation history. 

Oscillations due to self-inductance of the circuit, along with ohmic and radiation losses, are commonly used to resolve the paradox. In this text an attempt is made to isolate the inductance part in the form of wave-guiding phenomenon, providing yet another point of view of this intriguing educational problem. The treatment presented here stresses that any real capacitor must be of finite size which, apart from radiation, leads to a wave phenomenon inside the capacitor enabling it to
resolve the paradox by itself.

%%%%%%%%%%%%%%%%%%%%%%%%%%%%%%%%%%%%%%%%%%%%%%%%%%%%%%%%%%%%%%%%%%%%%%%%%%%%%%%%%%%%%%%%%%%%%%%%%%%%%%%%%%%
\section{Associated \ac{TRL} Problem}
\label{Sec:TRLProblem}

\begin{figure}
	\includegraphics[width=\figwidth cm]{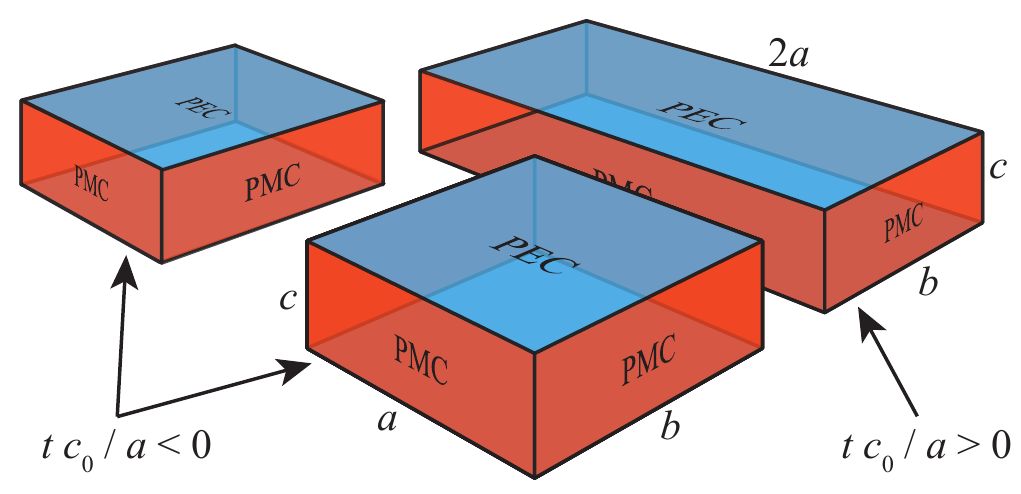}
	\caption{A sketch of the capacitor problem.}
	\label{fig:setup}
\end{figure}

To isolate the wave-guiding part of the capacitor problem, an idealized setup with two ideal vacuum-filled parallel plate capacitors is assumed, see Fig.~\ref{fig:setup}. Each capacitor is formed by a rectangular box of size $a \times b \times c$ which is bounded by two \ac{PEC} plates (size $a \times b$) and four \ac{PMC} plates (size $b \times c$ and $a \times c$). For normalized times $t c_0 / a < 0$, with $c_0$ being the speed of light, one of the capacitors has been charged to potential difference $U_0$ and the other has been left uncharged. At normalized time $t c_0 / a = 0$ the capacitors are put into contact, forming a larger capacitor of size $2a \times b \times c$ and the \ac{PMC} plate dividing it has been instantaneously removed.

The final setup for normalized times $t c_0 / a > 0$ can be recognized as an open-ended transmission line~\cite{Zangwill_Modern_Electrodynamics} in which a \ac{TEM} wave with voltage $U \left(\zeta,t\right)$ and current $I \left(\zeta,t\right)$ can propagate with $\zeta$ denoting a coordinate along dimension $a$, $\zeta \in \left(-a,a\right)$. The connection of capacitors can, in this setup, be seen as an initial condition
\begin{align}
\label{eq:init1}
\frac{U\left( {\zeta ,0} \right)}{U_0} &= \left| {\begin{array}{*{20}{c}}
	{1,\;\zeta \in \left( {0,a} \right)}\\
	{0 ,\;\zeta \in \left( { - a,0} \right)}
	\end{array}} \right.\\
\label{eq:init2}
\frac{Z_{\mathrm{TRL}} I\left( {\zeta ,0} \right)}{U_0} &= \left| {\begin{array}{*{20}{c}}
	{0,\;\zeta \in \left( {0,a} \right)}\\
	{0,\;\zeta \in \left( { - a,0} \right)}
	\end{array}} \right.
\end{align}
where $Z_{\mathrm{TRL}}$ denotes the characteristic impedance~\cite{Zangwill_Modern_Electrodynamics} of \ac{TRL}. The open ends further dictate boundary conditions
\begin{equation}
\label{eq:boundary}
\frac{Z_{\mathrm{TRL}} I\left( {-a ,t} \right)}{U_0} = \frac{Z_{\mathrm{TRL}} I\left( {a ,t} \right)}{U_0} = 0.
\end{equation}

%%%%%%%%%%%%%%%%%%%%%%%%%%%%%%%%%%%%%%%%%%%%%%%%%%%%%%%%%%%%%%%%%%%%%%%%%%%%%%%%%%%%%%%%%%%%%%%%%%%%%%%%%%%
\section{Solution to the \ac{TRL} Problem}
\label{Sec:TRLSolution}

The solution to the one-dimensional dissipationless problem stated in Sec.~\ref{Sec:TRLProblem} is given by current and voltage waves travelling along and against the $\zeta$ coordinate, specifically
\begin{align}
\label{eq:waves1}
U \left(\zeta,t\right) &= U^{+} \left(t - \frac{\zeta}{c_0} \right) + U^{-} \left(t + \frac{\zeta}{c_0}\right), \\
\label{eq:waves2}
Z_{\mathrm{TRL}} I \left(\zeta,t\right) &= U^{+} \left(t - \frac{\zeta}{c_0} \right) - U^{-} \left(t + \frac{\zeta}{c_0} \right).
\end{align}
To satisfy boundary conditions (\ref{eq:boundary}) it can be shown that functions $U^{+} \left(\tau \right)$ and $U^{-} \left(\tau \right)$ must be periodic functions with period $4a/c_0$ so that
\begin{equation}
\label{eq:periodic}
U^{+} \left( \tau - \frac{2a}{c_0} \right) = U^{-} \left( \tau \right) = \sum\limits_{n =  - \infty }^\infty  {f\left( \tau - \frac{4na}{c_0} \right)}, 
\end{equation}
with $f \left(\tau\right)$ being an arbitrary function defined over the normalized time interval of length $4$. The last step is to satisfy initial conditions. Using (\ref{eq:waves1}), (\ref{eq:waves2}) and (\ref{eq:periodic}), it can be checked that general initial conditions imply
\begin{align}
\zeta \in \left(-a,a\right) \\
f \left(\frac{\zeta}{c_0}\right) &= \frac{1}{2} \Big[ U \left(\zeta,0\right) - Z_{\mathrm{TRL}} I \left(\zeta,0\right) \Big] \\
\zeta \in \left(a,3a\right) \\
f \left(\frac{\zeta}{c_0} \right) &= \frac{1}{2} \Big[ U \left(-\zeta +2a,0\right) + Z_{\mathrm{TRL}} I \left(-\zeta +2a,0\right) \Big]
\end{align}
which uniquely defines periodic function (\ref{eq:periodic}).

For specific initial conditions (\ref{eq:init1}) and (\ref{eq:init2}), function $f$ is defined as
\begin{equation}
\label{eq:solution}
\frac{ \displaystyle f \left(\frac{\zeta}{c_0}\right)}{U_0}  = \left| {\begin{array}{*{20}{c}}
	{\displaystyle \frac{1}{2},\;\zeta \in \left( {0,2a} \right)}\\ [12pt] 
	{0 ,\; \mathrm{otherwise}}
	\end{array}} \right.
\end{equation}
Substituting (\ref{eq:solution}) into (\ref{eq:periodic}) and subsequently into (\ref{eq:waves1}) and (\ref{eq:waves2}) the resulting voltage and current waves are depicted in Fig.~\ref{fig:solution}.

\begin{figure}[t]
	\begin{center}
		\animategraphics[loop, controls, autoplay, buttonsize=0.75em,width=\figwidth cm]{1}{Animate_PDF/FigBB}{0}{32}
		\caption{Animation (available in Adobe Acrobat Reader) of normalized voltage and current waves after the capacitors have been connected. The initial frame shows the situation at normalized time $t c_0 / a = 0$ and the final frame shows the situation at normalized time $t c_0 / a = 4$. The depicted animation shows the time course during one period of the described wave phenomenon.}
		\label{fig:solution}
	\end{center}
\end{figure}

%%%%%%%%%%%%%%%%%%%%%%%%%%%%%%%%%%%%%%%%%%%%%%%%%%%%%%%%%%%%%%%%%%%%%%%%%%%%%%%%%%%%%%%%%%%%%%%%%%%%%%%%%%%%
\section{Discussion}
\label{disc}
This section discusses two noticeable issues concerning the simplistic approach shown in the previous section. The first is energy conservation, a topic that is at the very heart of the capacitor paradox problem~\cite{McDonald_capacitor_paradox}. The second issue touches to the ideal discontinuous waveforms shown in Fig.~\ref{fig:solution}.

\subsection{Energy Conservation}
\begin{figure}
	\includegraphics[width=\figwidth cm]{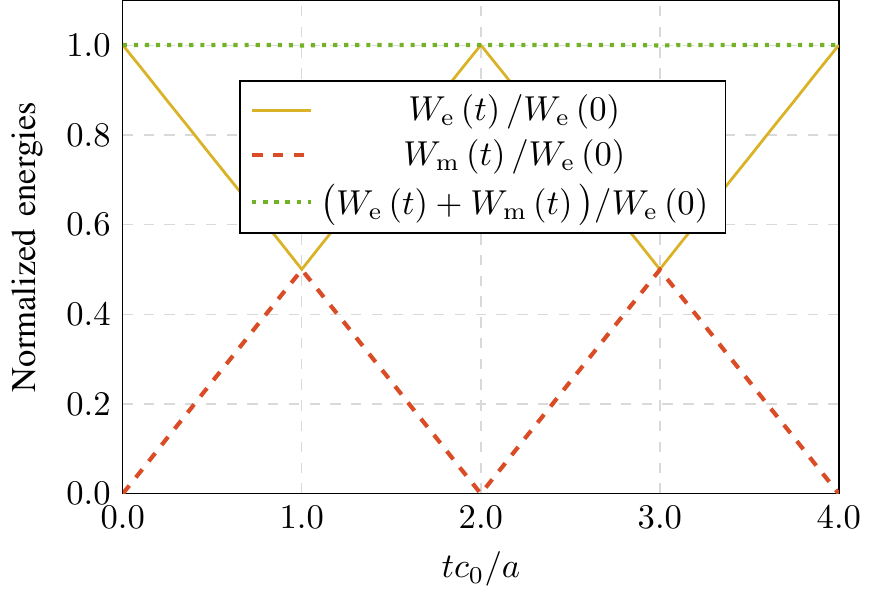}
	\caption{Electric and magnetic energies within the studied system in normalized time interval $t c_0 / a \in \left(0,4\right)$.}
	\label{fig:energies}
\end{figure}

Electric energy $W_{\mathrm{e}} \left(t\right)$ and magnetic energy $W_{\mathrm{m}} \left(t\right)$ within the studied system can be evaluated as~\cite{Zangwill_Modern_Electrodynamics}
\begin{align}
	\label{eq:WE}
	W_{\mathrm{e}} \left(t\right) &=\frac{1}{2} C_{\mathrm{pul}} \int\limits_{-a}^a \big| U \left(\zeta,t\right) \big|^2  \mathrm{d}\zeta , \\
	\label{eq:WM}
	W_{\mathrm{m}} \left(t\right) &=\frac{1}{2} C_{\mathrm{pul}} \int\limits_{-a}^a \big| Z_{\mathrm{TRL}} I \left(\zeta,t\right) \big|^2  \mathrm{d}\zeta,
\end{align}
with $C_{\mathrm{pul}}$ denoting per-unit-length capacitance. The time course of these energies for the current and voltage corresponding to Fig.~\ref{fig:solution} are shown in Fig.~\ref{fig:energies} and normalized with respect to initial energy $W_{\mathrm{e}} \left(0\right) = C_{\mathrm{pul}} a U_0^2 / 2$. It can be observed that at normalized time $t c_0 / a = 0$ the energy (purely electric) resides in one half of the structure, at normalized time $t c_0 / a = 2$ it resides in the second half of the structure and this repeats with period $t c_0 / a = 4$. Between these extremal times electric energy is transferred across the structure at the expense of magnetic energy generated by the underlying current flow. The total electromagnetic energy is conserved as it should be.

\subsection{Fringing fields}
The \ac{PMC} boundaries are used to enforce zero fringing fields at the edges of the capacitor. This greatly simplifies the solution, but also creates some interpretation difficulties. Before the removal of the \ac{PMC} diaphragm dividing the capacitors, their inner volumes have been electromagnetically isolated. It thus happens that just at the time of diaphragm's removal, there is a step discontinuous voltage, but current is still zero. The Telegrapher's equations~\cite{Zangwill_Modern_Electrodynamics}
\begin{align}
\label{eq:tele1}
- c_0 \frac{\partial U \left(\zeta,t\right)}{\partial \zeta} &=\frac{\partial Z_{\mathrm{TRL}} I \left(\zeta,t\right) }{\partial t}, \\
\label{eq:tele2}
- c_0 \frac{\partial Z_{\mathrm{TRL}} I \left(\zeta,t\right) }{\partial \zeta} &= \frac{\partial U \left(\zeta,t\right)}{\partial t},
\end{align}
then contains a dirac delta distribution on both sides and are satisfied only in a generalized sense. Another feature connected to this issue can be observed at normalized times $t c_0 / a = 1$ and $t c_0 / a = 3$ when there is a constant current across the structure which, by design, must drop to zero value at \mbox{$\zeta=-a,a$} during an infinitesimally short space interval.

In a real capacitor problem, there is a fringing field and thus, prior to the connection, the second capacitor is polarized by the fringing fields of the first one (and vice vesa). There is a current flow in the second capacitor even during the time when capacitors approach each other and the time evolution depends on how fast the capacitors have been connected. However, phenomenologically, voltage and current waves after connection roughly follow the time evolution described above, with smooth waveforms. Fringing fields further result in radiation which will dissipates energy.

%%%%%%%%%%%%%%%%%%%%%%%%%%%%%%%%%%%%%%%%%%%%%%%%%%%%%%%%%%%%%%%%%%%%%%%%%%%%%%%%%%%%%%%%%%%%%%%%%%%%%%%%%%%%
%\section{Conclusion}
%\label{concl}

%XXXXX


\begin{thebibliography}{1}
	\providecommand{\url}[1]{#1}
	\csname url@samestyle\endcsname
	\providecommand{\newblock}{\relax}
	\providecommand{\bibinfo}[2]{#2}
	\providecommand{\BIBentrySTDinterwordspacing}{\spaceskip=0pt\relax}
	\providecommand{\BIBentryALTinterwordstretchfactor}{4}
	\providecommand{\BIBentryALTinterwordspacing}{\spaceskip=\fontdimen2\font plus
		\BIBentryALTinterwordstretchfactor\fontdimen3\font minus
		\fontdimen4\font\relax}
	\providecommand{\BIBforeignlanguage}[2]{{%
			\expandafter\ifx\csname l@#1\endcsname\relax
			\typeout{** WARNING: IEEEtran.bst: No hyphenation pattern has been}%
			\typeout{** loaded for the language `#1'. Using the pattern for}%
			\typeout{** the default language instead.}%
			\else
			\language=\csname l@#1\endcsname
			\fi
			#2}}
	\providecommand{\BIBdecl}{\relax}
	\BIBdecl
	
	\bibitem{McDonald_capacitor_paradox}
	\BIBentryALTinterwordspacing
	K.~T. McDonald. (2002) A capacitor paradox. [Online]. Available:
	\url{http://www.hep.princeton.edu/{\texttildelow}mcdonald/examples/twocaps.pdf}
	\BIBentrySTDinterwordspacing
	
	\bibitem{Zangwill_Modern_Electrodynamics}
	A.~Zangwill, \emph{Modern Electrodynamics}.\hskip 1em plus 0.5em minus
	0.4em\relax Cambridge University Press, 2012.
	
\end{thebibliography}
\end{document}